\documentclass{article}
\usepackage{graphicx}
\usepackage{amsmath}

\begin{document}

\begin{center}
\bigskip {\Large Spin rotation and oscillation of high energy particles in
storage ring}

{\large V.G.Baryshevsky}

\emph{Research Institute of Nuclear Problems, Byelorussian State University,
Minsk 220050, Belarus}
\end{center}

The phenomenon of light polarization plane rotation (e.g., the Faraday
effect, the natural rotation of a light polarization plane) as well as light
birefringence (e.g., in matter placed in an electric field due to the Kerr
effect) are the well known optical coherent phenomena. For the first glance
they distinguish photons from other particles (nucleons, electrons, etc.)
for which these effects for a long time have been considered nonexistent.

In [1-8] a wide range of phenom-ena similar to the effects of light
polarization plane rotation and birefringence was shown to exist for
particles other than photons. In particular it has been shown that as
particles (neutrons, protons, neutrinos, etc.) pass through matter with
polarized nuclei, the particle spin undergoes a rotation in an effective
pseudomagnetic field of the matter induced by both strong and weak
interactions. This effect is kinematically analogous to the phenomenon of
light polarization plane rotation due to the Faraday effect. Experimentally,
the effect of neutron spin precession in polarized target has been studied
for neuterons [4-6].

As it was ascertained in [5-8] the analogue of birefringence phenomenon
exists for particles, too.

As a matter of fact, the Faraday effect and the ef-fect of birefringence are
caused by the dependence of the coherent photon-medium interaction energy on
the photon spin state. This property unites the quasioptical phenomena
discovered in [1-8] for interaction of spin-particles in matter with nuclei
with the phenomena existing in light optics. However, attention should be
drawn to the fact that whereas the photon spin is equal to unity, the
particle (atom, nucleus) spin may take on different values. For particles of
spin $S=1/2$ there exists only one effect - that of spin rotation, i.e. a
kinematic analog of the effect of light polarization plane rotation. An
effect similar to birefringence exists for spin $S\geq 1$ particles. It is
very interesting to mention that phenomena of rotation and oscillations of
particle spin (birefringence effect) exist for particles with spin $S\geq 1$
in a medium with unpolarized scatterer spins, too. The fact that these
effects are described by spin-dependent part of scattering amplitude allows
to use them for the measurement of this amplitude at different energies of
colliding particles.

\subsection{ Spin rotation of high-energy particles in polarized targets.}

\bigskip As a result of numerous studies (see, for example, \cite{12}), a
close connection between the coherent elastic scattering amplitude $f(0)$
and the refraction index of a medium $N$ has been established:

\begin{equation}
N=1+\frac{2\pi \rho }{k^{2}}f\left( 0\right)   \label{refr_ind}
\end{equation}
where $\rho $ is the number of particles per $cm^{3}$, $k$ is the wave
number of a particle incident on a target. In 1964 it was shown \cite{1rot}
that while slow neutrons are propagating through the target with polarized
nuclei a new effect of nucleon spin precession occured. It is stipulated by
the fact that in a polarized target the neutrons are characterized by two
refraction indices ($N_{\uparrow \uparrow }$ for neutrons with the spin
parallel to the target polarization vector and $N_{\uparrow \downarrow }$
for neutrons with the opposite spin orientation , $N_{\uparrow \uparrow
}\neq N_{\uparrow \downarrow })$. According to the \cite{2rot}, in the
target with polarized nuclei there is a nuclear pseudomagnetic field and the
interaction of an incident neutron with this field results in neutron spin
rotation. The results obtained in \cite{1rot}, initiated experiments which
proved the existence of this effect [9-11].

Thus, let us consider the amplitude of elastic coherent zero-angle
scattering of nucleon by polarized nucleon (nucleous).

The general form of this amplitude with allowance for strong electromagnetic
and weak interactions is given in \cite{2rot}. Below we shall consider more
concretely the effect of a relativistic nucleon spin rotation in the target
with polarized protons (nuclei with spin 1/2), caused by strong interaction.
In this case, the explicit structure of the elastic scattering amplitude \
of a particle with spin 1/2 by a particle with spin 1/2 \ (see, for example, 
\cite{13,14}) proceeds from the following simple discussions. In our case,
the elastic scattering amplitude at zero angle depends on the spin operators
\ $\frac{1}{2}\sigma $, $\frac{1}{2}\sigma _{1}$ of an incident particle \
and that of a target , and also on the momentum of the incident particle ,
that is, on $\overrightarrow{n}=\frac{\overrightarrow{k}}{k}$ . Operators $%
\overrightarrow{\sigma },$ $\overrightarrow{\sigma _{1}}$ can be contained
in the expression for the amplitude only in the first degree, as higher
degrees \ of $\overrightarrow{\sigma }$ reduce either to a number or to $%
\overrightarrow{\sigma }$. The combinations $\overrightarrow{\sigma },$ $%
\overrightarrow{\sigma _{1}}$ and $\overrightarrow{n}$ must be such that the
scattering amplitudes are \ a scalar and invariant in space and time
reflections. These conditions defenetely determine its general form:

\begin{equation}
\widehat{F}=A+A_{1}\left( \overrightarrow{\sigma }\cdot \overrightarrow{%
\sigma _{1}}\right) +A_{2}\left( \overrightarrow{\sigma }\cdot 
\overrightarrow{n}\right) \left( \overrightarrow{\sigma _{1}}\cdot 
\overrightarrow{n}\right) .  \label{amplitude}
\end{equation}
By averaging the amplitude $\widehat{F}$\ with the help of a spin matrix of
the density of scatters $\rho _{s}$ the elastic coherent scattering
amplitude may be written as:

\begin{equation}
f=Sp\,\rho _{s}\,\widehat{F}=A+A_{1}\left( \overrightarrow{\sigma }\cdot 
\overrightarrow{p}\right) +A_{2}\left( \overrightarrow{\sigma }\cdot 
\overrightarrow{n}\right) \left( \overrightarrow{n}\cdot \overrightarrow{p}%
\right)  \label{aver_amp}
\end{equation}
where $\overrightarrow{p}=Sp\,\rho _{s}$\thinspace $\overrightarrow{\sigma
_{1}}$ is the polarization vector of a scatterer in a target.

Amplitude $f$ can be expressed as

\begin{equation}
f=A+\overrightarrow{\sigma }\cdot \overrightarrow{g}  \label{ampl_g}
\end{equation}
where $\overrightarrow{g}=A_{1}\overrightarrow{p}+A_{2}\overrightarrow{n}%
\left( \overrightarrow{n}\cdot \overrightarrow{p}\right) $.

To simplify further reasoning let us consider the situation when vector $%
\overrightarrow{n}$ is either parallel to $\overrightarrow{p}$ \ ($%
\overrightarrow{n}\parallel \overrightarrow{p}$) or perpendicular to $%
\overrightarrow{p}$ ($\overrightarrow{n}\perp \overrightarrow{p}$).

In this case one has that $g\left( \overrightarrow{n}\parallel 
\overrightarrow{p}\right) =\left( A_{1}+A_{2}\right) \overrightarrow{p}$ and 
$g\left( \overrightarrow{n}\perp \overrightarrow{p}\right) =A_{1}%
\overrightarrow{p}$. Thus in these cases vector $\overrightarrow{g}$ is
directed along $\overrightarrow{p}.$ Selecting quantization axes parallel to 
$\overrightarrow{p}$, one can see that scattering amplitude $f_{\uparrow
\uparrow }=A+g$ \ of nucleon with spin parallel to $\overrightarrow{p}$ is
not equal to scattering amplitude $f_{\uparrow \downarrow }=A-g$ \ of
nucleon with spin antiparallel to $\overrightarrow{p}$. Hence, the
corresponding refractive indices are not equal to each other (i.e. $%
N_{\uparrow \uparrow }\neq N_{\uparrow \downarrow }$).

Considering a wave passes through a layer of polarized medium with finite
thickness one can express refractive index of nuclon with spin parallel to $%
\overrightarrow{p}$ as follows:

\begin{equation}
N_{\uparrow \uparrow }=1+\frac{2\pi \rho }{k^{2}}f_{\uparrow \uparrow }=1+%
\frac{2\pi \rho }{k^{2}}\left( A+g\right)  \label{n+}
\end{equation}

and for nucleon with opposite polarization

\begin{equation}
N_{\uparrow \downarrow }=1+\frac{2\pi \rho }{k^{2}}f_{\uparrow \downarrow
}=1+\frac{2\pi \rho }{k^{2}}\left( A-g\right)  \label{n-}
\end{equation}

then the difference

\begin{equation}
\Delta N=N_{\uparrow \uparrow }-N_{\uparrow \downarrow }=\frac{2\pi \rho }{%
k^{2}}\left( f_{\uparrow \uparrow }-f_{\uparrow \downarrow }\right) =\frac{%
4\pi \rho }{k^{2}}g  \label{dn}
\end{equation}

is determined by the difference in correspondent coherent scattering
amplitudes and differs from zero only in polarized medium.

Suppose that nucleon passes through polarized medium and their polarizations
are oriented at certain angle to the vector $\overrightarrow{p}$. This state
of nucleon can be described as superposition of two states with
polarizations directed along and opposite to the vector $\overrightarrow{p}$%
. Initial wave function of nucleon can be expressed as:

\begin{equation}
\psi \left( \overrightarrow{r}\right) =e^{i\overrightarrow{k}\overrightarrow{%
r}}\chi _{n},\,\,\chi _{n}=\left( 
\begin{array}{c}
c_{1} \\ 
c_{2}
\end{array}
\right)
\end{equation}

or

\begin{equation}
\psi \left( \overrightarrow{r}\right) =c_{1}e^{i\overrightarrow{k}%
\overrightarrow{r}}\left( 
\begin{array}{c}
1 \\ 
0
\end{array}
\right) +c_{2}e^{i\overrightarrow{k}\overrightarrow{r}}\left( 
\begin{array}{c}
0 \\ 
1
\end{array}
\right) .
\end{equation}

Suppose quantization axes z coincides with vector $\overrightarrow{p}$ and
particle falls onto the target orthogonally to its surface. As the state $%
\left( 
\begin{array}{c}
1 \\ 
0
\end{array}
\right) $ possesses refration index $N_{\uparrow \uparrow }$ and the state $%
\left( 
\begin{array}{c}
0 \\ 
1
\end{array}
\right) $ is characterized by $N_{\uparrow \downarrow }$, then the wave
function of nucleon in polarized medium changes with penetration depth as
follows:

\begin{equation}
\psi \left( \overrightarrow{r}\right) =\left( 
\begin{array}{c}
c_{1}\psi _{\uparrow \uparrow }\left( \overrightarrow{r}\right) \\ 
c_{2}\psi _{\uparrow \downarrow }\left( \overrightarrow{r}\right)
\end{array}
\right) =c_{1}\,\,e^{ikN_{\uparrow \uparrow }l}\left( 
\begin{array}{c}
1 \\ 
0
\end{array}
\right) +c_{2}\,\,e^{ikN_{\uparrow \downarrow }l}\left( 
\begin{array}{c}
0 \\ 
1
\end{array}
\right) ,.  \label{psi}
\end{equation}
$\ l$ is the pass length of nucleon in target.

Using (\ref{psi}) one can find nucleon polarization vector

\begin{equation}
\vec{p}_{n}=\langle \psi |\vec{\sigma}|\psi \rangle
\end{equation}
and as a result

\begin{eqnarray}
p_{nx} &=&2{Re}c_{1}^{\ast }c_{2}\psi _{+}^{\ast }\psi _{-}\langle \psi
|\psi \rangle ^{-1},
p_{ny}=2{Im}c_{1}^{\ast }c_{2}\psi _{+}^{\ast
}\psi _{-}\langle \psi |\psi \rangle ^{-1},  \label{p-project} \\
p_{nz} &=&\left( \left| c_{1}\psi _{+}\right| ^{2}-\left| c_{2}\psi
_{-}\right| ^{2}\right) \langle \psi |\psi \rangle ^{-1}.  \notag
\end{eqnarray}

Suppose that nucleon spin in vacuum is perpendicular to the polarization
vector of nuclei. Direction of nucleon spin in vacuum define as axes $x$. In
this case $c_{1}=c_{2}=1/\sqrt{2}.$ Using (\ref{p-project}) we obtain

\begin{eqnarray}
p_{nx} &=&\cos \left[ k{Re}\left( N_{\uparrow \uparrow }-N_{\uparrow
\downarrow }\right) \,l\right] \,e^{-k{Im}\left( N_{\uparrow \uparrow
}-N_{\uparrow \downarrow }\right) l}\langle \psi |\psi \rangle ^{-1},\, 
\notag \\
p_{ny} &=&-\sin \left[ k{Re}\left( N_{\uparrow \uparrow }-N_{\uparrow
\downarrow }\right) \,l\right] \,e^{-k{Im}\left( N_{\uparrow \uparrow
}-N_{\uparrow \downarrow }\right) l\,}\langle \psi |\psi \rangle ^{-1},
\label{p-2} \\
p_{nz} &=&\frac{1}{2}\left( \,e^{-2k{Im}N_{\uparrow \uparrow
}l\,}-e^{-2k{Im}N_{\uparrow \downarrow }l\,}\right) \langle \psi |\psi
\rangle ^{-1}=\frac{\,e^{-2k{Im}N_{\uparrow \uparrow }l\,}-e^{-2k{%
Im}N_{\uparrow \downarrow }l\,}}{e^{-2k{Im}N_{\uparrow \uparrow
}l\,}+e^{-2k{Im}N_{\uparrow \downarrow }l\,}}.  \notag
\end{eqnarray}
$\langle \psi |\psi \rangle =\frac{1}{2}\left( \,e^{-2k{Im}N_{\uparrow
\uparrow }l\,}+e^{-2k{Im}N_{\uparrow \downarrow }l\,}\right) $.

It should be reminded that $Im\,f(0)=\frac{k}{4\pi }\sigma _{tot}$, where $%
\sigma _{tot}$ is the total cross-section of scatterring of nucleon by
nucleon (nulei, atoms).

\bigskip According to (\ref{p-2}) when nucleon penetrate deep into target,
its polarisation vector rotates around nuclei polarization vector at the
angle

\begin{equation}
\vartheta =k{Re}\left( N_{\uparrow \uparrow }-N_{\uparrow \downarrow
}\right) l=\frac{2\pi \rho }{k}{Re}\left( f_{\uparrow \uparrow
}-f_{\uparrow \downarrow }\right) l  \label{theta}
\end{equation}

\bigskip This rotation is similar to the spin rotation appearing in magnetic
field. Then we can conclude that polarized nuclear target acts on spin
likewise the area occupied with nuclear pseudomagnetic field \cite{1rot}.

It is important to emphasize that in experiments with gas target scattering
amplitude $f(0)$, being contained in expression for $N$, is the elastic
coherent amplitude of zero-angle scattering of a nucleon by an atom (of
hydrogen, deuterium and so on). This fact should be taken into consideration
at particular analysis, because atom electrons can contribute to the
rotation angle.

Let us consider proton beam passing through polarized gas hydrogen target.
Let polarized atoms of hydrogen in magnetic field are described by the
magnetic quantum number $M=1$. It means that in hydrogen atom spins of
proton and electron are parallel. Polarized atoms of hydrogen possess
magnetic moment directed along spin, which produces magnetic field $%
B_{H}\sim \rho \,\mu _{H}$, where $\rho $ is the number of atoms of hydrogen
in $1\,cm^{3}$, $\mu _{H}$ is the magnetic moment of hydrogen atom. As the
magnetic moment of proton is much less than that of electron, then magnetic
field $B_{H}$ \ of polarized hydrogen atoms is mainly produced by their
electrons.

Hence the angle of rotation of proton spin can be expressed as a sum of two
additives $\vartheta _{B}$ and $\vartheta _{S}$

\begin{equation*}
\vartheta =\vartheta _{B}+\vartheta _{S},
\end{equation*}
where $\vartheta _{B}$\ is the angle of rotation caused by magnetic field
and $\vartheta _{S}$ is the angle of rotation appearing due to strong
interactions i.e. nuclear pseudomagnetic field.

Thus, considering rotation phenomena as a tool for experimental
investigation of \ zero-angle nucleon-nucleon scattering amplitude one come
to neccessity to extract addition caused by magnetic field of the target $%
\overrightarrow{B}=\overrightarrow{H}+\overrightarrow{B_{H}}$ , here $%
\overrightarrow{H}$ is the external magnetic field. This can be done:

1. by calculation using equations of Bargman-Michel-Telegdi (BMT)\emph{\ }
\cite{14} as magnetic moments of proton and electron (hydrogen atom) are
known with high accuracy

or

2. by experimental separating of additions caused by magnetic and nuclear
pseudomagnetic fields. This possibility is due to the fact that magnetic
field induced by polarized magnetic moment depends on the shape of the
target because of long-distance action of electromagnetic interaction.
Whereas the nuclear pseudomagnetic field (being short-range action) does not
depend on the shape of the target.

Moreover, from the analysis of BMT-equations follows that if proton velocity
is directed along $\overrightarrow{B}$\ then angle of rotation of proton
spin orthogonal to $\overrightarrow{B}$\ $\ $for nonrelativistic protons is
determined by magnetic moment. If \ proton velocity is orthogonal to $%
\overrightarrow{B}$\ than spin rotation in magnetic field is determined by
anomalous magnetic moment, since due to cyclotron movement of proton in
magnetic field the addition to the angle of rotation, caused by Dirac\
magnetic moment \ (equal to the nuclear magneton), yields to the
conservation of angle between proton spin and momentum (if the anomalous
magnetic moment $\Delta \mu $ is equal to zero). Hence the observed
deviation of proton spin in magnetic field $\overrightarrow{B}$\textbf{\ }
is conditioned by $\Delta \mu $.

Let us evaluate the effect magnitude for particular setup \cite{15}.
According to \cite{15} target thickness is $n=\rho l=10^{14}\,atoms/cm^{2}$,
a revolution frequency of proton beam $\nu \sim 10^{6}\,s^{-1}.$

From (\ref{theta}) we can obtain for rotation angle caused by nuclear
pseudomagnetic field :

\begin{eqnarray*}
\vartheta _{S} &=&\frac{2\pi \rho l}{k}{Re}\left( f_{\uparrow \uparrow
}^{NN}(0)-f_{\uparrow \downarrow }^{NN}(0)\right) \cdot \nu T=\frac{4\pi
\,n\nu T}{k}{Re}g= \\
&=&\frac{4\pi n\nu T}{\sqrt{\frac{2ME}{\hbar ^{2}}}}{Re}g=\frac{4\pi
\hbar n\nu {Re}gT}{\sqrt{2ME}}\cong 1\,rad
\end{eqnarray*}
$T$ is the observation time. (for example, $T=10\,\,hours=3.6\cdot 10^{4}\,s$%
, $g\cong 2\cdot 10^{-13}\,cm)$

It is interesting to note that we can obtain simple estimation, showing
relation between angle of proton spin rotation caused by magnetic field $%
\overrightarrow{B_{e}}$ , produced by polarized magnetic moment of electrons
(atoms), \ and that in nuclear pseudomagnetic field. It should be reminded
that $B_{e}=\eta 4\pi \rho \mu _{e}$, where $\eta $ depends on the target
shape (for example, for sphere $\eta =\frac{2}{3}$) \cite{Jackson}.

Suppose polarization vector $\overrightarrow{p}$ is orthogonal to the
particle momentum. In this case the addition to the rotation angle caused by 
$B_{e}$ is expressed 
\begin{equation*}
\vartheta _{Be}=\frac{k\Delta \mu _{p}B_{e}l}{E}=\eta \frac{k\Delta \mu
_{p}\,4\pi \rho \mu _{e}l}{E},
\end{equation*}
where $\Delta \mu _{p}$ is the anomalous magnetic moment of proton , $l$ is
the length of gas target , $E$ is the particle energy.

Then we obtain

\begin{equation*}
\frac{\vartheta _{Be}}{\vartheta _{s}}=\eta \frac{\Delta \mu _{p}\,\mu _{e}}{%
g\frac{\hbar ^{2}}{2M}}\cong 0.9\,\eta \,\frac{r_{0}}{g},
\end{equation*}
where $r_{0}=\frac{e^{2}}{mc^{2}}$ is the electromagnetic radius of
electron. And for $g=2\cdot 10^{-13}$ and$\,\,r_{0}=2.8\cdot 10^{-13}$ \
ratio $\dfrac{\vartheta _{Be}}{\vartheta _{s}}\cong \eta .$

\subsection{ Deuteron spin rotation and oscillations in a nonpolarized target%
}

According to the above, the refractive index of neutral and charged
particles of spin $S$ can be written as

\begin{equation}
\widehat{N}=1+\frac{2\pi \rho }{k^{2}}\widehat{f}\left( 0\right)
\label{ampl_hat}
\end{equation}
where $f(0)$ is the amplitude of particle zero-angle elastic coherent
scattering by a scattering center, which is an operator acting in the
particle spin space, $\widehat{f}\left( 0\right) =Sp\,\widehat{\rho }_{J}$%
\thinspace $\widehat{F}\left( 0\right) $ ; $\,\widehat{\rho }_{J}$ is the
spin density matrix of the scatterer; $\widehat{F}\left( 0\right) $ stands
for the forward scattering operator amplitude acting in the spin space of
the particle and the scatterer of spin $\overrightarrow{J}$. If the wave
function of the particle entering a target is $\psi _{0}$, then that after
travelling a distance $z$ in the target is written as

\begin{equation}
\psi \left( z\right) =\exp \left( i\,k\,\widehat{N}\,z\right) \psi _{0}
\label{wave_func}
\end{equation}

\bigskip The explicit form of the amplitude $\widehat{f}\left( 0\right) $
for particles with arbitrary spin $S$ \ has been obtained in \cite{6rot}.
According to these articles even for an unpolarized target is a function of
the incident particle spin operator and can be written as

\begin{equation}
\widehat{f}\left( 0\right)
=d+d_{1}S_{z}^{2}+d_{2}S_{z}^{4}+...+d_{s}S_{z}^{2s}  \label{f_hat}
\end{equation}

The quantization axis $z$ is directed along $\vec{n}=\frac{\overrightarrow{k}%
}{k}$. Consider a specific case of strong interactions invariant under space
and time reflections. For this reason, the terms containing odd powers of $S$
are neglected. Correspondingly, the refractive index

\begin{equation}
\widehat{N}=1+\frac{2\pi \rho }{k^{2}}\left(
d+d_{1}S_{z}^{2}+d_{2}S_{z}^{4}+...+d_{s}S_{z}^{2s}\right)  \label{N_hat}
\end{equation}

From eq. (\ref{N_hat}) one can draw an important conclusion about the
refractive index being dependent on the spin orientation with respect to the
pulse direction. Let $m$ denote a magnetic quantum number, then for a
particle in a state that is an eigenstate of the spin projection operator
onto the $z$ axis, $S_{z}$, the refractive index is written as

\begin{equation}
N\left( m\right) =1+\frac{2\pi \rho }{k^{2}}\left(
d+d_{1}m^{2}+d_{2}m^{4}+...+d_{s}m^{2s}\right)  \label{N_m}
\end{equation}

According to eq. (\ref{N_m}), the particle states with quantum numbers $m$
and $-m$ have the same refractive indices. For a spin-1 particle (for
example, a $J/\psi $ particle, a deuteron) and a spin-$\frac{3}{2}$ particle
(i.g. an $\Omega ^{-}$ hyperon). 
\begin{equation*}
N\left( m\right) =1+\frac{2\pi \rho }{k^{2}}\left( d+d_{1}m^{2}\right)
\end{equation*}

As seen,

\begin{eqnarray*}
{Re}\,N\left( \pm 1\right) &\neq &{Re}\,N\left( 0\right) , \\
{Im}\,N\left( \pm 1\right) &\neq &{Im}\,N\left( 0\right) , \\
{Re}\,N\left( \pm \frac{3}{2}\right) &\neq &{Re}\,N\left( \pm 
\frac{1}{2}\right) \\
{Im}\,N\left( \pm \frac{3}{2}\right) &\neq &{Im}\,N\left( \pm 
\frac{1}{2}\right)
\end{eqnarray*}

Since we have obtained the explicit spin structure of the refractive index
and the wave function (\ref{wave_func}) is known, in every specific case we
can find all spin properties of a beam in the target at depth $z$.

Let us come to consideration of deuteron passing through medium (particle
with spin = 1). The wave function can be represented as a superposition of
basis spin wave functions $\chi _{m}$, which are eigenfunctions of the
operators $\widehat{S}^{2}$ and $\widehat{S}_{z},$ $\widehat{S}_{z}\chi
_{m}=m\chi _{m}$:

\begin{equation*}
\psi =\sum_{m=\pm 1,0}a^{m}\chi _{m}.
\end{equation*}
Let us look for the mean value $\langle \vec{S}\rangle =\langle \psi \left| 
\widehat{\vec{S}}\right| \psi \rangle /\langle \psi |\psi \rangle $ of the
spin operator in state $\psi .$

Suppose the particle enter the target at $z=0$. Wave function of the
particle insid the medium at the depth z can be expressed as:

\begin{equation*}
\Psi =\left\{ 
\begin{array}{c}
a^{1} \\ 
a^{0} \\ 
a^{-1}
\end{array}
\right\} =\left\{ 
\begin{array}{c}
a\,e^{i\delta _{1}}e^{ikN_{1}z} \\ 
b\,e^{i\delta _{0}}e^{ikN_{0}z} \\ 
c\,e^{i\delta _{-1}}e^{ikN_{-1}z}
\end{array}
\right\} =\left\{ 
\begin{array}{c}
a\,e^{i\delta _{1}}e^{ikN_{1}z} \\ 
b\,e^{i\delta _{0}}e^{ikN_{0}z} \\ 
c\,e^{i\delta _{-1}}e^{ikN_{1}z}
\end{array}
\right\}
\end{equation*}
it should be mentioned that $N_{1}=N_{-1}$

\bigskip Let us choose coordinate system in which plane $\left( xz\right) $
coincides with that formed by vector $\langle \overrightarrow{S}\rangle $
(\bigskip $<\vec{S}\mathbf{>=}\frac{<{\psi }\mathbf{|}\vec{S}\mathbf{|}{\psi 
}>}{\mid \psi \mid ^{2}})$\ before entering the target and deuteron
momentum. In this case $\delta _{1}-\delta _{0}=\delta _{-1}-\delta _{0}=0$
and components of vector at $z=0$ $\,\,\,\langle S_{x}\rangle \neq 0,\langle
S_{y}\rangle =0.$

\bigskip\ As a result we obtain:

\begin{eqnarray}
&<&{S}_{x}>=\sqrt{2}{e^{-\frac{1}{2}\rho (\sigma _{0}+\sigma _{1})z}b(}{a+c)}%
\cos [\frac{2\pi \rho }{k}{Re}d_{1}\,z]/|\psi |^{2},  \notag \\
&<&{S}_{y}>=-\sqrt{2}{e^{-\frac{1}{2}\rho (\sigma _{0}+\sigma _{1})z}b(}{a-%
\emph{c})}\sin [\frac{2\pi \rho }{k}{Re}d_{1}z]/|\psi |^{2},  \label{S_}
\\
&<&{S}_{z}>={e^{\rho \sigma _{1}z\,}(}{a}^{2}-\emph{c}^{2}{)}/|\psi |^{2}, 
\notag
\end{eqnarray}

Particle with spin 1 also possesses tensor polarization i.e. tensor of
quadrupolarization \textbf{\ $\hat{Q}_{ij}=3/2(\hat{S}_{i}\hat{S}_{j}+\hat{S}%
_{j}\hat{S}_{i}-4/3\delta _{ij})$ }

for it we can obtain

\begin{eqnarray}
&<&{Q}_{xx}>=\left\{ -\left[ a^{2}+c^{2}\right] \frac{1}{2}\,e^{-\rho \sigma
_{1}\,z}+{b}^{2}\,e^{-\rho \sigma _{0}\,z}+3e^{-\rho \sigma
_{1}\,z}\,ac\,\cos \left[ \delta _{1}-\delta _{-1}\right] \right\} /|\psi
|^{2}  \notag \\
&<&{Q}_{yy}>=\left\{ -\left[ a^{2}+c^{2}\right] \frac{1}{2}\,e^{-\rho \sigma
_{1}\,z}+{b}^{2}\,e^{-\rho \sigma _{0}\,z}-3e^{-\rho \sigma
_{1}\,z}\,ac\,\cos \left[ \delta _{1}-\delta _{-1}\right] \right\} /|\psi
|^{2}  \notag \\
&<&{Q}_{zz}>=\left\{ \left[ a^{2}+c^{2}\right] \frac{1}{2}\,e^{-\rho \sigma
_{1}\,z}-2{b}^{2}\,e^{-\rho \sigma _{0}\,z}\right\} /|\psi |^{2}~,
\label{quadr} \\
&<&{Q}_{xy}>=-3e^{-\rho \sigma _{1}\,z}\,ac\,\sin \left[ \delta _{1}-\delta
_{-1}\right] /|\psi |^{2},  \notag \\
&<&{Q}_{xz}>=\frac{3}{\sqrt{2}}{e^{-\frac{1}{2}\rho (\sigma _{0}+\sigma
_{1})z}b(}{a-\emph{c})}\cos [\frac{2\pi \rho }{k}{Re}d_{1}z]/|\psi
|^{2},  \notag \\
&<&{Q}_{yz}>=-~\frac{3}{\sqrt{2}}{e^{-\frac{1}{2}\rho (\sigma _{0}+\sigma
_{1})z}b(}{a+\emph{c})}\sin [\frac{2\pi \rho }{k}{Re}d_{1}z]/|\psi
|^{2}~,  \notag
\end{eqnarray}
where $|\psi |^{2}=2{}(a^{2}+c^{2})\,e^{-\rho \sigma
_{1}\,z}+{}b^{2}\,e^{-\rho \sigma _{0}\,z}$ and $z$ is the length of
particle path inside a medium.

\bigskip According to (\ref{S_},\ref{quadr}) rotation appears if angle
between polarization vector and momentum of particle differs from $\frac{\pi 
}{2}$. At this for acute angle between polarization vector and momentum the
sign of rotation is opposite than that for obtuse angles.

If spin is orthogonal to momentum then $(a=c)$ particle spin (tensor of
quadrupolarization) oscillate (do not rotate)

\begin{eqnarray}
&<&{S}_{x}>=\sqrt{2}{e^{-\frac{1}{2}\rho (\sigma _{0}+\sigma _{1})z}}%
2{}{}\cos [\frac{2\pi \rho }{k}{Re}d_{1}z]/|\psi |^{2},  \notag \\
&<&{S}_{y}>=0,  \notag \\
&<&{S}_{z}>=0,  \notag
\end{eqnarray}

And tensor of quadrupolarization:

\begin{eqnarray}
&<&{Q}_{xx}>=\biggl\{2{}^{2}\,e^{-\rho \sigma _{1}\,z}+{}^{2}\,e^{-\rho
\sigma _{0}\,z}\}/|\psi |^{2}~,  \notag \\
&<&{Q}_{yy}>=\biggl\{-4{}^{2}\,e^{-\rho \sigma _{1}\,z}/2+{}^{2}\,e^{-\rho
\sigma _{0}\,z}\biggr\}/|\psi |^{2}~,  \notag \\
&<&{Q}_{zz}>=\biggl\{2{}^{2}\,e^{-\rho \sigma _{1}\,z}-2\,{}^{2}\,e^{-\rho
\sigma _{0}\,z}\biggr\}/|\psi |^{2}~,  \label{quad} \\
&<&{Q}_{xy}>=0,  \notag \\
&<&{Q}_{xz}>=0,  \notag \\
&<&{Q}_{yz}>=\biggl\{-\frac{3}{\sqrt{2}}{e}^{\left( \sigma _{0}+\sigma
_{1}\right) z}\,{2}{}ab\,\sin [\frac{2\pi \rho }{k}{Re}d_{1}z]\biggr\}%
/|\psi |^{2}~~,  \notag
\end{eqnarray}
where $|\psi |^{2}=2{}^{2}\,e^{-\rho \sigma _{1}\,z}+{}^{2}\,e^{-\rho \sigma
_{0}\,z}$ and $z$ is the length of particle path inside a medium.

\bigskip Let us evaluate phase of oscillation

\begin{equation*}
\varphi =\frac{2\pi \rho \;z}{k}{Re}d_{1}
\end{equation*}
for COSY. In this case total phase storing during experiment $\Phi =\varphi
\nu T$. For particular conditions $\rho \;z=n=10^{14}cm^{-2}$ , $Red_{1}\sim
10^{-13}$ (for scattering of deuteron by hydrogen)$,\nu \sim 10^{6},T\sim
70\,\,hours=2.52\cdot 10^{5\,}\,s$ and $\Phi $ can be estimated as $\Phi
\sim 1\,rad$.

As you can see the effect magnitude is large enough to be observable at
COSY. It is very important that in considered case (deuterons passing
through nonpolarized medium) there are no magnetic fields at the target area.

\bigskip Experimental study of $\ f\left( 0\right) $ at different particle
energies is an important tool for investigation of particles interaction
properties. Phenomena of spin rotation and oscillation, described above,
give the basis for methods of investigation of amplitude $\ f\left( 0\right) 
$ spin-dependendent part, examination of dispesion relations for it and
testing of P-,T-violations at particle interactions for different energies.
Of the essence, interactions at final state make the measurement of T-odd
contributions difficult at experimental studies of T-violation by scattering
of particles by each other. This problem is absent for T-odd particle spin
rotation due to refraction, because it is determined by the elastic coherent
zero-angle scattering amplitude. In this case initial and final states of
target coincides and the above difficulty does not occur. As a result,
methods based on the measurement of spin rotation angle (and $f\left(
0\right) $ measurement, as a sequence) can provide the most tight
restrictions to the possible value of T-odd interactions.

One more promising aria of experimental investigations can be mentined,
where phenomena, caused by nuclear optics of polarized medium, can appear
efficient. Storage rings become now the more and more important
investigation tool. Life-time of particle beam \ in storage ring can reach
several hours and even days. During this time particles circulate in storage
ring with frequency of several MHz and even small spin-dependent
interactions of beams with each other can significantly influence on
polarization state of beams.

We considered the rest target above. But in storage rings moving bunch\ is
usually used as a target. No problems appear at study of spin rotation
effect in this case - you should consider the effect in the rest frame of
one of the beams and then reduce the result to the laboratory frame.

As an example let us consider cross-collision of two bunches of polarized
particles. Suppose particles of the first beam have mass $m_{1}$, energy $%
E_{1}$and Lorentz factor $\gamma _{1}$, whereas particles of the second beam
are characterized by mass $m_{2}$, energy $E_{2}$ and Lorentz factor $\gamma
_{2}$. Choose the frame where the second beam rest. In this frame the energy
of the particles of the first beam $E_{1}^{\,\,\prime }=E_{1}\gamma
_{2}=m_{1}c^{2}\gamma _{1}\gamma _{2}$. Refractive index is expressed in
conventional form:

\begin{equation*}
\widehat{N}=1+\delta \widehat{N}+\frac{2\pi \rho _{2}^{\prime }}{%
k_{1}^{\prime 2}}\widehat{f}\left( E_{1}^{\,\,\prime },0\right)
\end{equation*}
where $\delta \widehat{N}$ is the contribution to refractive index caused by
refraction of particles in electric and magnetic fields of the bunch, $\rho
_{2}^{\prime }=\gamma _{2}^{-1}\rho _{2}$ is the density of bunch 2 in its
rest frame and $\rho _{2}$ is the density of the second bunch in laboratory
frame, $k_{1}^{\prime }=k_{1}\gamma _{2}$. is the wave number of particles
of the first bunch in the rest frame of bunch 2 and $k_{1}$ is the wave
number of these particles in laboratory frame. Contribution to the particle
spin rotation angle caused by $\widehat{f}\left( E_{1}^{\,\,\prime
},0\right) $ is expressed:

\begin{equation*}
\vartheta ^{\prime }=\frac{2\pi \rho _{2}^{\prime }}{k_{1}^{\prime }}\left( 
{Re}\,f_{\uparrow \uparrow }\left( E_{1}^{\,\,\prime },0\right) -{%
Re}\,f_{\uparrow \downarrow }\left( E_{1}^{\,\,\prime },0\right) \right) L,
\end{equation*}
where $L=\gamma _{2}l$ is the length of the bunch 2 in its rest frame, $l$
is the length of this bunch in laboratory frame. Lorentz factor of particle
1 in rest frame of particle 2 is $\gamma =\gamma _{1}\gamma _{2}$ and this
factor can be educed in scattering amplitude: $f\left( E_{1}^{\,\,\prime
},0\right) =\gamma _{1}\gamma _{2}\,f^{\prime }\left( E_{1}^{\,\,\prime
},0\right) $. As a result:

\begin{equation*}
\vartheta ^{\prime }=2\pi \rho _{2}\lambda _{1c}\left( {Re}%
\,f\,\,_{\uparrow \uparrow }^{\prime }\left( E_{1}^{\,\,\prime },0\right) -%
{Re}\,f\,\,_{\uparrow \downarrow }^{\prime }\left( E_{1}^{\,\,\prime
},0\right) \right) l,
\end{equation*}
where $\lambda _{1c}=h/m_{1}c$.

\bigskip Angle of rotation is invariant. Then in laboratory frame it will be
the same. Evidently, particles 2 also experience refraction on the bunch 1.
As a result mutual rotation of polarizatuion vectors of both bunches occurs.
Taking into consideration the fact that in storage ring particle passes
through the target \ (bunch) many times one can see that total angle of spin
rotation, caused by refraction, many times rises. The above makes
measurement of zero-angle scattering amplitude in wide particle energy range
very promising.

Very interesting possibility appears when one of the beams (e.g. beam 2) has
very low energy. It can be beam of ultracold atoms. Then index of refraction
of beam 2 by beam 1 is high and we can apply atom interferometry methods for
scattering amplitude measurement.

\bigskip It should be mentioned that unique possuibility to measure
spin-independent zero-angle scattering amplitude at high energies\ also
appears.

It is important to note that photon beam formed by laser wave can be used as
one of the beams. In this case effect of spin rotation is determined by
elastic coherent amplitude of zero-angle scattering of particle by photon.
According to the above analysis both electromagnetic and P-,T- odd weak
interactions contribute to the refration index of particle in the area
occupied by photons \cite{17}. As a result area occupied by photons can be
described as optically anisotropic medium.

\bigskip 

\end{document}